\documentclass[amsmath,amssymb,a4paper,notitlepage,floatfix]{revtex4-1}
\usepackage{graphicx}
\usepackage[paperwidth=212mm,paperheight=297mm,centering,hmargin=2.8cm,vmargin=2.6cm]{geometry}

\usepackage{physics}
\usepackage{chemformula}
\usepackage{float}
\usepackage[separate-uncertainty = true,multi-part-units=single]{siunitx}
\DeclareSIUnit{\molar}{M}
\DeclareSIUnit{\counts}{cnt}
\usepackage{jabbrv}

\begin{document}

\title{Polymer-encapsulated organic nanocrystals for single photon emission}

\author{Ross C. Schofield, Dominika P. Bogusz, Rowan A. Hoggarth, Salahuddin Nur, Kyle D. Major, and Alex S. Clark}
\email{alex.clark@imperial.ac.uk}
\affiliation{Centre for Cold Matter, Blackett Laboratory, Imperial College London, Prince Consort Road, SW7 2AZ London, United Kingdom}

\date{\today}
\begin{abstract}
We demonstrate an emulsion-polymerisation technique to embed dibenzoterrylene-doped anthracene nanocrystals in polymethyl methacrylate (PMMA) nanocapsules. The nanocapsules require no further protection after fabrication and are resistant to sublimation compared to unprotected anthracene. The room temperature emission from single dibenzoterrylene molecules is stable and when cooled to cryogenic temperatures we see no change in their excellent optical properties compared to existing growth methods. These now robust nanocapsules have potential for surface functionalisation and integration into nanophotonic devices, where the materials used are compatible with incorporation in polymer-based designs.
\end{abstract}
\let\newpage\relax
\maketitle
\section{Introduction}

Organic molecules can be bright, efficient sources of photons and when cooled to cryogenic temperature demonstrate enviable properties \cite{Kozankiewicz2014}. Specifically, polycyclic aromatic hydrocarbons encased in organic crystals have been shown to generate photons of high purity that show potential for use in applications that exploit the single photon nature of their emission \cite{Tian2014,Siyushev2014,Rattenbacher2019}.  These molecules can be processed at ambient temperature \cite{Pazzagli2018} and can be readily incorporated into photonic structures \cite{Turschmann2017,Grandi2019,Ciancico2019}. Many of the desirable properties, including ease of manufacture, are due to weak van der Waals bonding in aromatic crystals, which while useful can lead to experimental difficulties. The crystals have a substantial vapour pressure at room temperature and pressure, and therefore readily sublime away. Recent demonstration of self-assembled nanocrystals of organic molecules show ideal properties at low temperature, but if left unprotected have limited room temperature application \cite{Pazzagli2018}. We show here that by protecting self-assembled nanocrystals in a polymer shell we retain the ideal properties and allow these encapsulated nanocrystals to survive a wider variety of manufacturing processes and temperatures.

The exemplary molecule dibenzoterrylene (DBT) embedded in an anthracene (Ac) host matrix shows many of the characteristics of an ideal emitter. Initial work used co-sublimation methods \cite{Nicolet2007,Trebbia2009,Trebbia2010,Major2015} to produce macroscopic crystals, which show excellent spectral properties. However, to achieve the high collection efficiency required for applications the emitters need to be incorporated in nanophotonic devices. This would be greatly facilitated by introducing molecules into smaller crystals of the host material to minimise undesirable effects. Previous work has relied on other methods of incorporation. Melt-growth has been successful \cite{Faez2014,Turschmann2017,Rattenbacher2019}, but is has also been shown that this can be detrimental to the emission properties of the molecule due to confined growth \cite{Gmeiner2016a}. A super-saturated vapour growth is useful for making thin films on a surface \cite{Polisseni2016,Wei2019}, but so far demonstrations using this method have only been at room temperature.
The self-assembled precipitation method shown by Pazzagli \emph{et al.} required protection in a polymer layer which can also modify the nanophotonic environment introducing, for example, scattering loss \cite{Pazzagli2018}.

Our approach, detailed in this paper, is to protect the nanocrystals while still in solution to create polymethyl methacrylate (PMMA) encapsulated DBT containing Ac nanocrystals, or nanocapsules. These are produced through an emulsion polymerisation method \cite{Ahangaran2019}, where methyl methacrylate (MMA) is polymerised in solution around Ac nanocrystals. These PMMA nanocapsules show stable and spectrally-narrow single photon emission without further polymer coating or manipulation. The polymer nanocapsules also show robust physical properties, surviving long term storage at ambient temperature, and increased resistance to sublimation at elevated temperature. We believe the lack of additional polymer coating required and their physical robustness make them ideal for integration into nanophotonic devices. This approach has proved fruitful for inorganic quantum dot structures and the method can also allow functionalisation of the polymer coating \cite{Gao2008}. This can be useful for locating molecules at desirable positions on nanostructures and for biological applications where site labeling is needed. This method can also be easily extended for similar organic systems over a wide range of emission wavelengths \cite{Tamarat2000}.

\section{Synthesis and Experimental Setup}

\begin{figure}
    \centering
    \includegraphics[width=0.8\columnwidth]{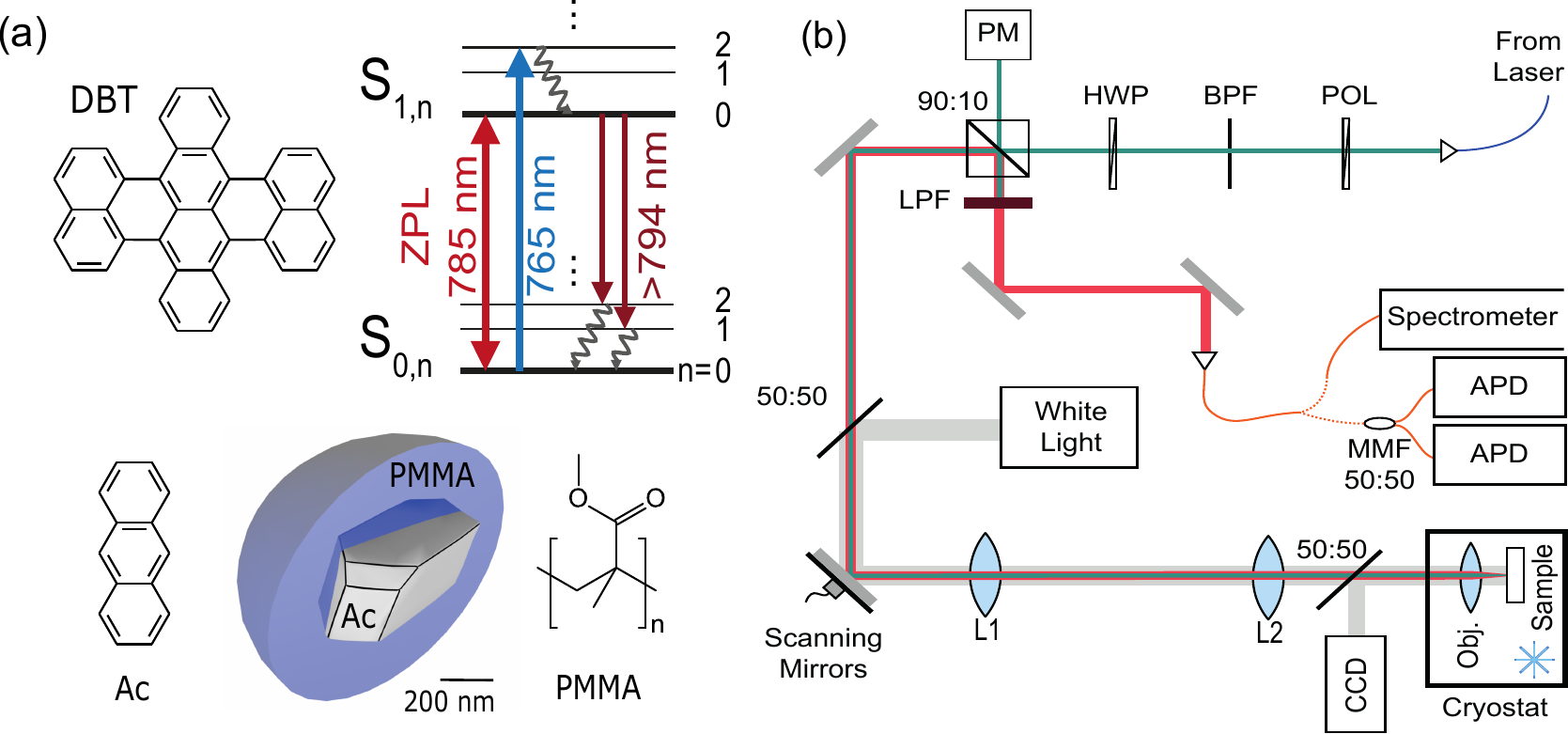}
    \caption{(a)~Chemical structures of DBT, Ac and PMMA, and an energy level diagram of DBT. The two electronic energy levels of interest, $S_{0,n}$ and $S_{1,n}$, are shown, along with their vibration sub-levels. The zero-phonon line (ZPL, \SI{784}{\nano\meter}) is shown as a transition between the $S_{0,0}$ and $S_{1,0}$ states. Blue-detuned pumping to a higher vibrational excited state (\SI{765}{\nano\meter}), fast non-radiative decay and red shifted fluorescence (>\SI{794}{\nano\meter}) are also shown. A diagram of a cut through of a PMMA nanocapsule (blue) containing an Ac nanocrystal (grey) is shown below. (b)~A schematic of the confocal microscope used for the fluorescence imaging of the nanocapsules. Dark green beam indicates the pump light, red is the fluorescence and grey is the white light used for imaging. POL:~linear polariser; BPF: band-pass filter; HWP: half-wave plate; 90:10:  90\% reflection: 10\% transmission beamsplitter; PM: power meter;  50:50: 50\% reflection: 50\% transmission beamsplitter; L1: first lens; L2: second lens; Obj: microscope objective lens; CCD:~charge-coupled device camera;  LPF: long-pass filter; MMF 50:50: multimode fibre beam splitter; APD: avalanche photodiode.}
    \label{fig:setup}
\end{figure}

A two-step growth process was used for synthesis to ensure nanocrystal/polymer shell distinction, resulting in a stable DBT environment and optimal fluorescence properties. The nanocrystals were synthesised first \cite{Pazzagli2018}. \SI{20}{\milli\litre} of distilled water was deoxygenated by sonication for 5 minutes. \SI{5}{\micro\litre} of \SI{1}{\micro\mol} DBT (MercaChem) in toluene (VWR) solution was added to \SI{10}{\milli\litre} of \SI{5}{\milli\mol} zone-refined anthracene (Tokyo Chemical Industry UK) in acetone (VWR) solution. \SI{2}{\milli\litre} of this mixed solution was then added to the distilled water and sonicated at \SI{37}{\kilo\hertz} and \SI{50}{\degreeCelsius} for 30~minutes. The vial was left uncovered for the growth. 

The PMMA encapsulation was then performed using an emulsion polymerisation method based on recipes used for the encapsulation of simple hydrocarbon cores \cite{Alkan2009,Sari2009,Sari2010,Alay2011}. \SI{2}{\micro\litre} of a 10:1 methyl methacrylate and allyl methacrylate solution (both Sigma-Aldrich, with no inhibitor and used as supplied), \SI{7}{\micro\litre} of \SI{10}{\micro\mol} ammonium persulfate solution (\ch{(NH_4)_2S_2O_8}, Sigma-Aldrich) and \SI{0.1}{\micro\gram} of ferrous sulphate heptahydrate (\ch{FeSO4 *}7\ch{H2O}, Sigma-Aldrich) were added to the solution and sonicated for 10~minutes whilst the sonicator heated to \SI{80}{\degreeCelsius}. The allyl methacrylate,  ammonium persulfate and ferrous sulphate heptahydrate are added with the MMA to increase the uniformity of formed particles and achieve a smooth, regular surface \cite{Sari2009}. There is also no additional surfactant added as the low concentration of reactants and the vigorous sonication mean it is not required.

Once the MMA monomer and Ac nanocrystals have formed an emulsion the polymerisation is initiated. \SI{10}{\micro\litre} of \SI{10}{\micro\mol} sodium thiosulphate solution (\ch{Na_2S_2O_7}, Sigma-Aldrich) and \SI{10}{\micro\litre} of 70\% tert-butylhydroperoxide (\ch{C4H10O2}, Sigma-Aldrich) were added. The solution was sonicated at \SI{80}{\degreeCelsius} for a further 2~hours to form the PMMA nanocapsules. The MMA preferentially polymerises around the Ac nanocrystals due to the decrease in solubility that occurs as polymer chain length increases. This decreased solubility makes it energetically favourable to form around the also immiscible Ac nanocrystals \cite{Ahangaran2019}. Fig.~\ref{fig:setup}(a) shows a diagram of a nanocapsule denoting the core/shell structure achieved.

Once the sonication solution had cooled $\sim$\SI{5}{\micro\litre} were pipetted onto a \SI{10}{\milli\meter} by \SI{10}{\milli\meter} substrate of 120\,nm \ch{Si_3N_4} coated on top of \SI{1}{\micro\meter} \ch{SiO_2} on a \SI{500}{\micro\meter} thick \ch{Si} wafer for analysis. This substrate was used to ensure good thermal conductivity when placed into our cryogenic confocal microscope. Prior to this, white light microscopy was performed using a Nikon Eclipse L200. 

Confocal fluorescence microscopy was performed using the confocal microscope shown in Fig.~\ref{fig:setup}(b). A tunable continuous wave Ti:Sapphire laser (SolSTiS, MSquared) or pulsed Ti:Sapphire laser (Tsunami, Spectra) was coupled into the microscope and collimated. The beam passed through a polariser (POL) to clean the polarisation and then a band-pass filter (BPF) to reduce background. A half-wave plate (HWP) was used to align the polarisation of the beam with the molecule dipole. A few percent was picked off here and sent to a photodiode for power-locking. A 90:10  beamsplitter was used in transmission with the reflected beam incident on a power meter (PM). The beam was then steered using electronically controlled galvanometer mirrors (Thorlabs, Scanning Mirrors) before passing through a `4-f' configuration lens setup (L1 and L2) into the back aperture of an objective lens (LD EC Epiplan-Neofluor 100x, 0.75~NA, Zeiss, Obj.). The sample was mounted after the objective on a 3-axis piezo positioning stage (Attocube). The objective lens and sample were mounted within a cryostat (Cryostation, Montana Instruments). This `4-f' configuration setup allowed raster-scanning of the \SI{720}{\nano\meter} full-width half-maximum focal spot across the sample via adjustment of the angles of the galvanometer mirrors. 

Fluorescence from the sample passed back through the shared beam path to the 90:10 beamsplitter, where the reflected beam was long-pass filtered (LPF) to remove the excitation laser light before being coupled to a multimode fibre. This was either sent to an avalanche photodiode (APD, Count-T, Laser Components), a multimode fiber beamsplitter (MMF 50:50) connected to two APDs for correlation measurements, or a spectrometer (Shamrock 303i with Newton EMCCD, Andor). A timing card (Hydraharp, PicoQuant) was used for any time-tagging and correlation experiments.

Two removable 50:50 beamsplitters, one before the galvanometer mirrors to shine in white light and the second between the second `4-f' lens and the objective lens to direct light to a CCD (iXon, Andor), were used for white light imaging.

\section{Results}
\begin{figure}
    \centering
    \includegraphics[width=0.8\columnwidth]{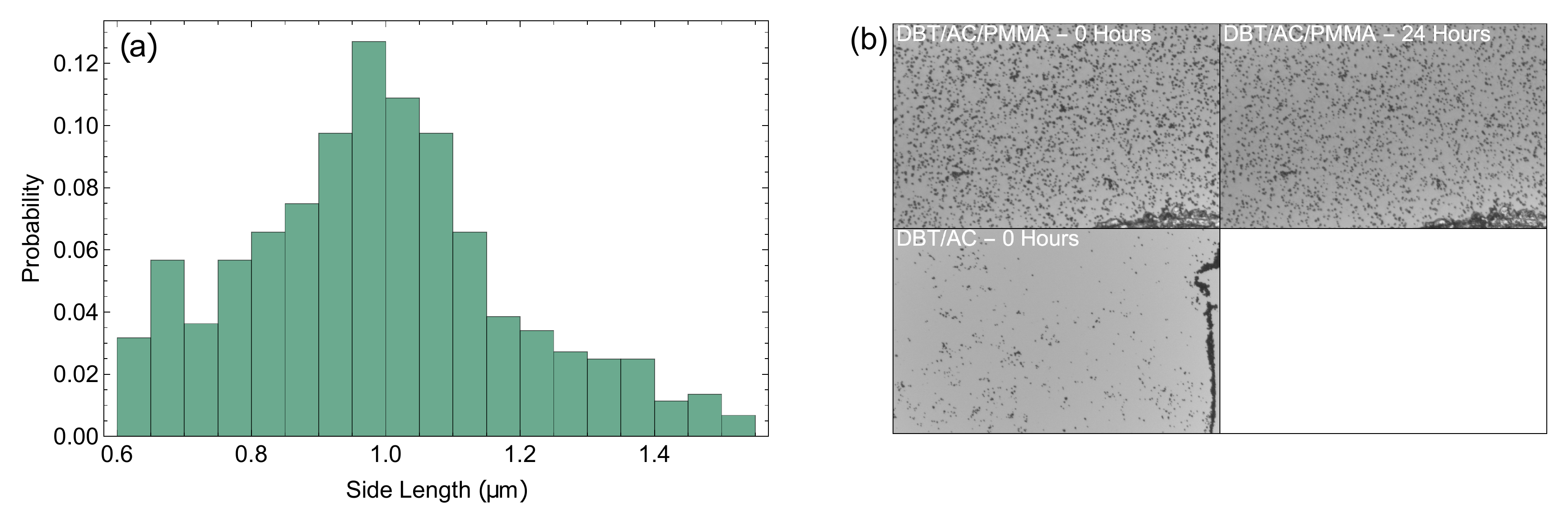}
    \caption{(a)~Histogram of single nanocapsule diameters taken from a 50x magnification image. (b)~White light images (20x magnification) of PMMA nanocapsules after deposition (top left) and 24 hours later (top right) and nanocrystals after deposition (bottom left) and 24 hours later (bottom right). Loss of material due to sublimation is clearly visible.}
    \label{fig:image}
\end{figure}

We first determined the presence of polymer nanocapsules by white light microscopy. Fig.~\ref{fig:image}(a) shows a histogram of nanocapsule size determined from a 50x magnification image. A mean nanocapsule size of \SI{990\pm50}{\nano\meter} with a standard deviation of \SI{200\pm50}{\nano\meter} is found. Growth times from 1 to 4 hours were investigated, with no visible change occurring beyond 2 hours.

One advantage of this encapsulation is preventing sublimation. To test this, images of polymer nanocapsules and uncoated nanocrystals taken 24 hours apart were compared. Fig.~\ref{fig:image}(b) shows these images. Analysis shows $\sim\SI{96}{\percent}$ of the nanocapsules remain, compared to $\sim\SI{10}{\percent}$ of the nanocrystals. This highlights the resistance to sublimation of the polymer nanocapsules provides, and implies the polymer shell fully encases the Ac.

A nanocapsule sample was then placed into the confocal microscope to investigate the properties of the DBT molecules contained within. This was first performed at room temperature. White light imaging with a CCD camera was used to identify PMMA nanocapsules. Laser light at a wavelength of 767~nm was then spatially raster scanned across the sample whilst measuring red-shifted photons on an APD to identify DBT containing nanocapsules. The blue-detuned 767~nm light is capable of exciting the broadened $S_{0,0}\rightarrow S_{1,n}$ transitions in DBT, as shown in Fig.~\ref{fig:setup}(a). Fast ps-timescale non-radiative relaxation then occurs from the excited vibrational electronic excited state $S_{1,n}$ to the ground vibrational electronic excited state $S_{1,0}$, where the red-shifted ns-timescale radiative decay to the ground state manifold $S_{0,n}$ can be detected using an APD~\cite{Schofield2018}. 

\begin{figure*}
    \centering
    \includegraphics[width=0.85\columnwidth]{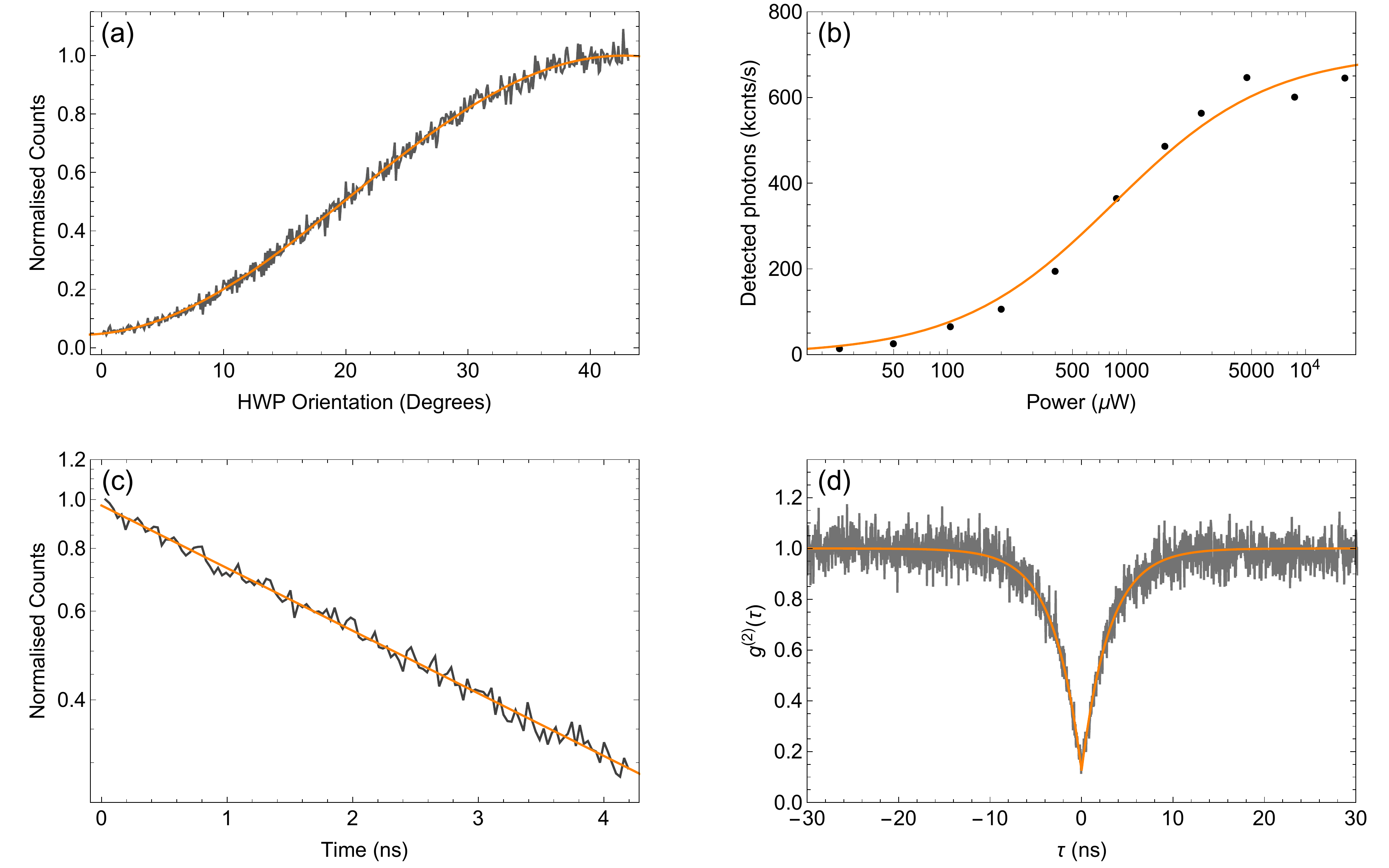}
    \caption{(a)~Black is normalised count rate from the selected molecule as the half-wave plate is rotated, fitted a $\sin^2$ curve shown in orange. (b)~Data points are the count rate measured from fitting a Gaussian to a spatial scan of the molecule at increasing pump powers. Fitted line is of the form given in Eq.~\ref{eq:sat}, giving a maximum count rate of \SI{710(20)}{\kilo\counts\per\second}. (c)~Black is data from a TCSPC measurement. Orange line is an exponential decay fitted to the tail of the data giving a lifetime of \SI{3.6(2)}{\nano\second}. (d)~The black is data from the second-order correlation function measurement exciting with $P=\SI{1.26(1)}{\milli\watt}$, showing a characteristic photon anti-bunching dip. Orange fitted line is of the form given in the text, showing a visibility of \SI{88(3)}{\percent}. }
    \label{fig:rt}
\end{figure*}

A nanocapsule containing a single DBT molecule was then selected for further analysis. The polarisation of the laser beam was varied to optimise the overlap with the molecular transition dipole. The resulting change in fluorescence is shown in Fig.~\ref{fig:rt}(a). Confocal scans at increasing laser intensity were then performed and fitted with 2D Gaussians~\cite{Schofield2018}. The maximum count rate at each intensity was extracted from this and fitted with a saturation curve, shown in Fig.~\ref{fig:rt}(b), using the function
\begin{equation}
\label{eq:sat}
R=R_{\infty}\frac{S}{1+S} \, ,
\end{equation}
where $R$ is the detected count rate, $R_{\infty}$ is the maximum count rate and $S=P/P_\text{sat}$ is the saturation parameter for the molecule with the laser power used, $P$, and the power at saturation, $P_\text{sat}$. From the data we find $R_\infty = \SI{710(20)}{\kilo\counts\per\second}$ and $P_\text{sat}$ = \SI{900(100)}{\micro\watt}.
The excited state lifetime of the emitter was measured using a time-correlated single photon counting experiment. The molecule was excited with a pulsed laser and the exponential decay in fluorescence with detection time is shown in Fig.~\ref{fig:rt}(c). Fitting an exponential decay to this data we find an excited state lifetime of $\tau_{1}=3.6$\,ns, in line with what is expected from DBT in Ac nanocrystals \cite{Pazzagli2018}. To confirm the quantum statistics of the emitted light, a second-order correlation function, $g^{(2)}(\tau)$, measurement was performed by splitting the emitted light onto two detectors and measuring the time delay $\tau$ between coincident detection. The normalised data, shown in Fig.~\ref{fig:rt}(d), was fitted using the function
\begin{equation}
g^{(2)}(\tau)=1 - \mathcal{V} e^{-(S+1)\Gamma_1|\tau|} \, ,
\end{equation}
where $\Gamma_1 = 1/\tau_{1}$ is the decay rate of the $S_{1,0}\rightarrow S_{0,n}$ transition and $\mathcal{V}$ is the visibility of the anti-bunching dip. Using the measured value for $\Gamma_1$, a visibility of $\mathcal{V}=$~\SI{88\pm3}{\percent} and a saturation parameter of $S=$~\SI{0.17(6)} are found. This measurement clearly shows the expected anti-bunching behaviour, and $\mathcal{V}$ agrees with the predicted value of $\mathcal{V}=$~\SI{85(4)}{\percent} from the signal to background ratio pumping at this power \cite{Lombardi2018}. Similarly, the $S$ value matches the expected value from the laser power used and the previous saturation measurements of $S=0.16\pm0.01$.  

These measurements show no significant deviation from expected values for DBT in both nanocrystals \cite{Pazzagli2018} and larger co-sublimation grown crystals \cite{Schofield2018}. Additionally, this molecule was excited at high intensities for many hours without any blinking or change in spectral properties, highlighting the suitability of the nano-encapsulation for protecting the emitter.

For any generation of coherent, indistinguishable photons from DBT the thermal dephasing from vibrations in the surrounding Ac matrix needs to be reduced. To do this we prepared a sample with higher DBT doping by increasing the volume of DBT:toluene solution from \SI{5}{\micro\litre} to \SI{10}{\micro\litre} during nanocrystal growth, and cooled the resulting encapsulated nanocrystals down to 4.7~K in a Montana Cryostation cryostat. The same confocal microscope was then used to analyse the sample. 

\begin{figure*}
    \centering
    \includegraphics[width=0.85\columnwidth]{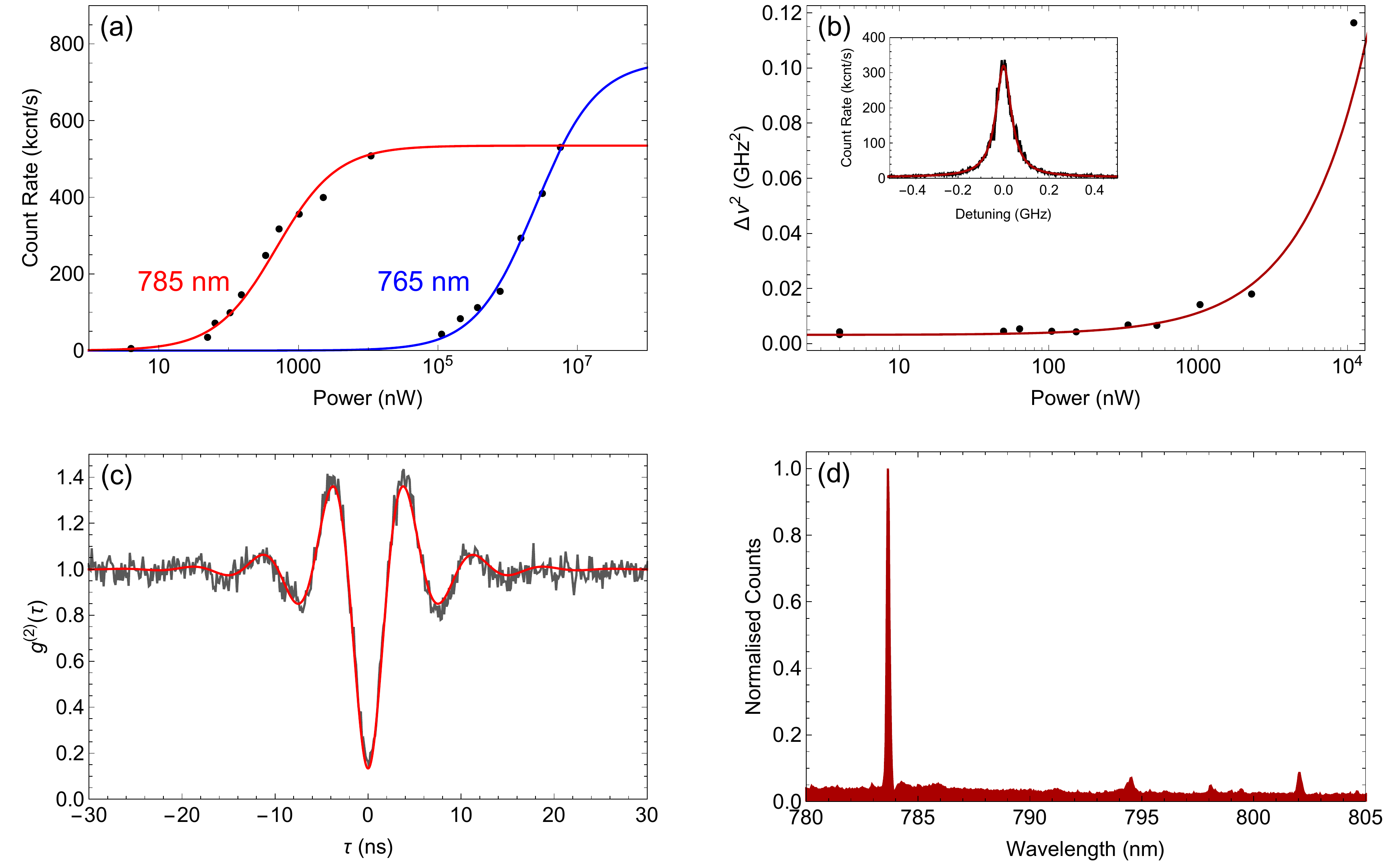}
    \caption{(a)~Count rates for resonant (\SI{784}{\nano\meter}, points with red curve) and blue-detuned (\SI{765}{\nano\meter}, points with blue curve) pumping for different pump powers. Resonant data is taken from fitted linescans (see~(b)) and blue-detuned data is from fitting a Gaussian to spatial fluorescence scans. Fitted lines are of the form given in Eq.~\ref{eq:sat}, giving $R_\infty=$~\SI{540(40)}{\kilo\counts\per\second} and $R_\infty=$~\SI{760(50)}{\kilo\counts\per\second} for the resonant and blue-detuned cases respectively. (b)~Data points are linewidth as a function of pump power taken from fitting Lorentzians to linescans. Fitted line is of the form given in Eq.~\ref{eq:pb}. Inset:~Example of detected fluorescence as a function of excitation wavelength showing the Lorentzian shape of a molecule resonance, see Eq.~\ref{eq:lorentz}, exciting with $P=\SI{530(1)}{\nano\watt}$. (c)~The black is data from the second-order correlation function measurement exciting with $P=\SI{5.49(1)}{\micro\watt}$, showing a characteristic photon anti-bunching dip. Red line is of the form given in Eq.~\ref{eq:g2full}. (d)~Fluorescence spectrum of a single molecule.}
    \label{fig:cold}
\end{figure*}

The laser was set to a wavelength of 784\,nm and a power $P=$~\SI{1}{\milli\watt}, then scanned across the sample to locate DBT containing nanocapsules through classical scattering and changes in background counts. The laser power was then reduced to $P=$~\SI{1}{\micro\watt}, expected to be $S\sim10$ \cite{Clear2020}, and scanned in wavelength from 780~nm to 785~nm while focused on a nanocapsule. During the scan we monitored the red-shifted fluorescence at wavelengths $>800$\,nm, collected and detected on an APD. When the laser was resonant with the $S_{0,0}\rightarrow S_{1,0}$ transition of a molecule a Lorentzian response was seen on the APD, as shown in the inset of Fig.~\ref{fig:cold}(b). This allowed us to identify a number of molecules in each nanocapsule.

After locating a DBT molecule, resonant linescans were repeated at increasing powers. The lineshape is described by the Lorentzian curve
\begin{equation}
\label{eq:lorentz}
R=R_{\infty}\frac{S}{1+S+\left(\tfrac{2\pi\delta}{\Gamma_2}\right)^2} \, ,
\end{equation}
where $\delta = \nu - \nu_0$ is the detuning of the laser frequency, $\nu$, from the molecule transition frequency, $\nu_0$, and $\Gamma_2$ is the dephasing rate of the transition \cite{Loudon2000}. The peak height of the Lorentzian can be plotted versus power and fitted with Eq.~\ref{eq:sat} to give a saturation curve, as shown in Fig.~\ref{fig:cold}(a). Fitting this data we find $R_\infty = \SI{540(40)}{\kilo\counts\per\second}$ and $P_\text{sat}$ = \SI{480(70)}{\nano\watt}.

A property of interest is the dephasing rate $\Gamma_2$, which is the inverse of the coherence time. Many quantum information uses require all emitted photons to be indistinguishable from one another, a property which is maximised when $\Gamma_2$ is reduced to lifetime-limit of $\Gamma_2=\Gamma_1/2$ \cite{Clear2020,Rezai2018}. The measured linewidth $\Delta\nu$ at a given excitation power is related to $\Gamma_2$ through \cite{Grandi2016}
\begin{equation}
\label{eq:pb}
    \Delta\nu=\tfrac{1}{\pi}\Gamma_2\sqrt{1+S} \, ,
\end{equation}
as shown in Fig.~\ref{fig:cold}(b). The saturation dependence here is due to the phenomenon of power broadening. A `zero-power' linewidth of \SI{56\pm4}{\mega\hertz} is found, corresponding to $\Gamma_2=2\pi~\times~\SI{28\pm2}{\mega\hertz}$.

The second order correlation function was then measured. The resulting data is shown in Fig.~\ref{fig:cold}(c). Due to the greatly reduced thermal dephasing at low temperatures, the $g^{(2)}(\tau)$ function takes the form \cite{Grandi2016}
\begin{equation}
\label{eq:g2full}
g^{(2)}(\tau)=1-\mathcal{V}\frac{p+q}{2q}e^{-\frac{1}{2}(p-q)|\tau|}+\mathcal{V}\frac{p-q}{2q}e^{-\frac{1}{2}(p+q)|\tau|}\, ,
\end{equation}
where $p=\Gamma_1 +\Gamma_{2}$, $q=\sqrt{(\Gamma_1-\Gamma_{2})^2-4\Omega^2}$, and $\Omega = \sqrt{\Gamma_1\Gamma_{2}S}$.

Coherent Rabi oscillations are seen either side of the characteristic anti-bunching dip. $\Gamma_2$ and $P_\text{sat}$ are known from the previous measurements of linewidth and saturation, and the fit gives a visibility $\mathcal{V} = $~\SI{90\pm10}{\percent} and a decay rate of $\Gamma_1=2\pi \times \SI{46\pm6}{\mega\hertz}$. The predicted visibility is $\mathcal{V} =$~\SI{96\pm2}{\percent}, when taking into account background from the pump laser and photons from other nearby molecules. This background could be reduced by using a lower concentration of DBT during the nanocapsule synthesis \cite{Clear2020}, thereby increasing the measured visibility. Our measured linewidth is broader than the lifetime-limit ($2\pi \times \Delta\nu=2\Gamma_2=\Gamma_1$) by a factor $\sim1.2$ due to excess thermal dephasing, consistent with what is expected for the temperature of our cryostat \cite{Clear2020}.

The laser was then tuned to 765~nm to excite an $S_{0,0}\rightarrow S_{1,n}$ transition, similar to previous room-temperature experiments. Spatial laser scans were performed, whilst recording the \SI{785\pm3}{\nano\meter} band-pass filtered fluorescence to maximise signal-to-background. This data was fitted with a saturating Gaussian to determine the saturation properties of the molecule, shown by the blue curve in Fig.~\ref{fig:cold}(a). We find $R_\infty = \SI{760(50)}{\kilo\counts\per\second}$ and $P_\text{sat}$ = \SI{2500(400)}{\micro\watt} on fitting this data. The increase in $P_\text{sat}$ compared to the case of resonant pumping is mainly due to the \SI{40}{\giga\hertz} width of the vibrational transition, which is $\sim$1000 times broader than the $S_{0,0}\rightarrow S_{1,0}$ transition \cite{Schofield2018}. To maximise the signal-to-background we adjusted the detuning of the laser from the vibrational transition to avoid exciting other nearby molecules. This detuning caused a further increase in the observed value of $P_\text{sat}$. The spectrum shown in Fig.~\ref{fig:cold}(d) was measured on the spectrometer whilst pumping the $S_{0,0}\rightarrow S_{1,3}$ transition. The spectrum shows the strong $S_{1,0}\rightarrow S_{0,0}$ zero-phonon line (ZPL) at \SI{783.7}{\nano\meter}, and the first three $S_{1,0}\rightarrow S_{0,n\neq0}$ decays to vibrational levels at \SI{794.5}{\nano\meter}, \SI{798.1}{\nano\meter} and \SI{802.0}{\nano\meter}. These are at typical wavelengths for DBT in Ac \cite{Deperasinska2011,Clear2020}.

\begin{figure}
    \centering
    \includegraphics[width=0.85\columnwidth]{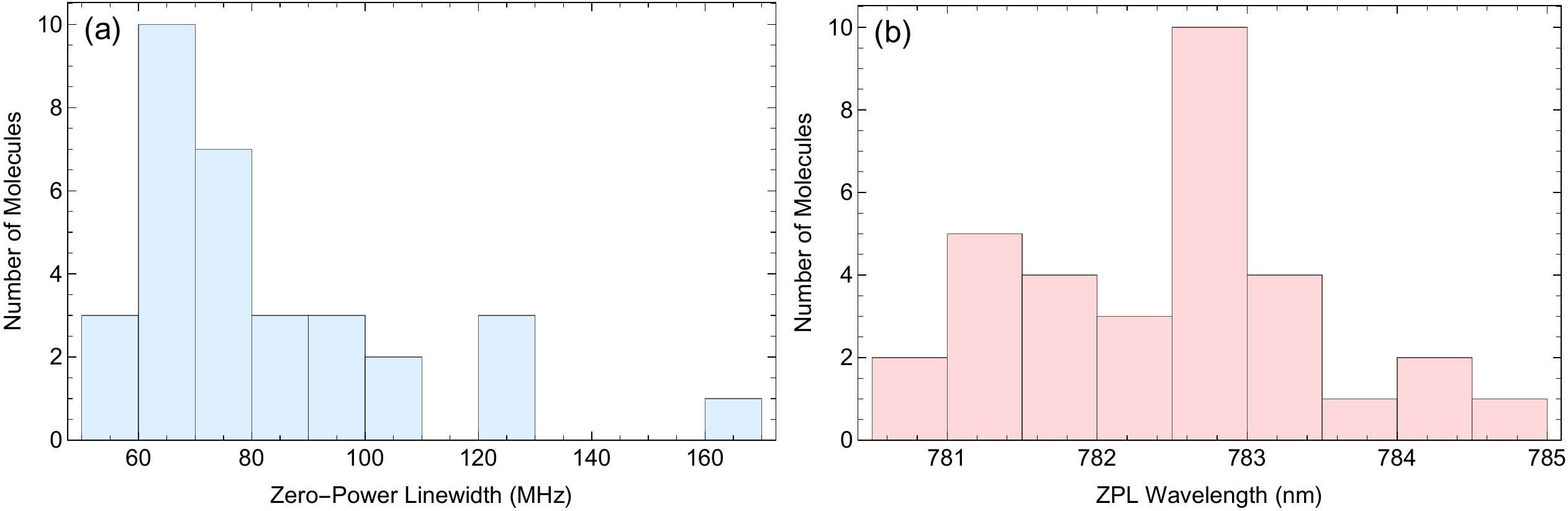}
    \caption{(a) Zero-power linewidth and (b) ZPL wavelength distributions for 35 DBT molecules in nanocapsules.}
    \label{fig:manymols}
\end{figure}

This data shows the excellent low temperature spectral properties of DBT are not affected by the PMMA encapsulation for this molecule. In order to determine if this molecule was a typical example, resonant linescans at a range of powers were performed for 34 additional molecules. Linewidths taken from these scans were fitted with Eq.~\ref{eq:pb} to give the zero-power linewidth. Fig.~\ref{fig:manymols}(a) shows a histogram of this data, showing many narrow, stable molecules. This large collection of molecules with linewidths around the limit of what is achievable with our cryostat shows the reliability of this nanoencapsulation method for single photon generation. Fig.~\ref{fig:manymols}(b) shows the central wavelength of the ZPL transition for these molecules, showing the inhomogeneous broadening due to local environmental differences and is consistent with bulk values.



\section{Conclusion}
This work shows the development of a self-assembled DBT/Ac/PMMA nanocapsules for single photon generation. The nanocapsules do not require any further treatment to survive continuous high-power excitation and ambient conditions for many hours and can emit near lifetime-limited photons at low temperature. The DBT shows no change in optical properties compared to already established growth methods, such as re-precipitation \cite{Pazzagli2018}, melt-growth \cite{Rattenbacher2019} and co-sublimation \cite{Major2015} showing there are no adverse effects due to this encapsulation method. 

We believe the lack of need for any further protection makes them ideal for incorporation into nanophotonic structures. Polymer based nanophotonic structures would be ideal candidates for integration due the similar materials used and the nanocapsules can now be subject to a wider range of manufacturing processes \cite{Knauer2017,Stella2019}. Microscopy shows a distribution of nanocapsule size with a mean of \SI{990\pm50}{\nano\meter}, however previous work has shown promise using commercially available syringe pore-filters to select smaller nanocrystals \cite{Pazzagli2018,Clear2020}, and this could be used for greater size selectivity of nanocapsules.

PMMA is a versatile material and functionalisation could be achieved through addition of modified monomers during the polymerisation step of the fabrication. Other work with PMMA encapsulation has already shown how the capsule properties, such as thickness and surface morphology, can be modified \cite{Ahangaran2019}. This works extends the practicality of organic emitters in quantum technologies, opening up new potential applications for these ideal sources of light.

\section*{Funding}
UKRI EPSRC (EP/P030130/1, EP/P01058X/1, EP/R044031/1, EP/P510257/1, and EP/L016524/1).
The Royal Society (UF160475).
EraNET Cofund Initiative QuantERA under the European Union Horizon 2020 research and innovation programme, Grant No. 731473 (ORQUID).

\section*{Acknowledgments}
We thank Ed Hinds for stimulating discussions, Jon Dyne and Dave Pitman for their expert mechanical workshop support, and Victoria Clark for graphic design work. 

\section*{Disclosures}
The authors declare no conflicts of interest.

\bibliographystyle{osajnl}
\bibliography{library}

\begin{thebibliography}{10}
\newcommand{\enquote}[1]{``#1''}

\bibitem{Kozankiewicz2014}
B.~Kozankiewicz and M.~Orrit, \enquote{{Single-molecule photophysics, from
  cryogenic to ambient conditions},} {\protect\JournalTitle{Chemical Society
  Reviews}} \textbf{43}, 1029--1043 (2014).

\bibitem{Tian2014}
Y.~Tian, P.~Navarro, and M.~Orrit, \enquote{{Single molecule as a local
  acoustic detector for mechanical oscillators},}
  {\protect\JournalTitle{Physical Review Letters}} \textbf{113}, 135505 (2014).

\bibitem{Siyushev2014}
P.~Siyushev, G.~Stein, J.~Wrachtrup, and I.~Gerhardt, \enquote{{Molecular
  photons interfaced with alkali atoms},} {\protect\JournalTitle{Nature}}
  \textbf{508}, 66--70 (2014).

\bibitem{Rattenbacher2019}
D.~Rattenbacher, A.~Shkarin, J.~Renger, T.~Utikal, S.~G{\"{o}}tzinger, and
  V.~Sandoghdar, \enquote{{Coherent coupling of single molecules to on-chip
  ring resonators},} {\protect\JournalTitle{New Journal of Physics}}
  \textbf{21}, 062002 (2019).

\bibitem{Pazzagli2018}
S.~Pazzagli, P.~Lombardi, D.~Martella, M.~Colautti, B.~Tiribilli, F.~S.
  Cataliotti, and C.~Toninelli, \enquote{{Self-Assembled Nanocrystals of
  Polycyclic Aromatic Hydrocarbons Show Photostable Single-Photon Emission},}
  {\protect\JournalTitle{ACS Nano}} \textbf{12}, 4295--4303 (2018).

\bibitem{Turschmann2017}
P.~T{\"{u}}rschmann, N.~Rotenberg, J.~Renger, I.~Harder, O.~Lohse, T.~Utikal,
  S.~G{\"{o}}tzinger, and V.~Sandoghdar, \enquote{{Chip-based all-optical
  control of single molecules coherently coupled to a nanoguide},}
  {\protect\JournalTitle{Nano Letters}} \textbf{17}, 4941--4945 (2017).

\bibitem{Grandi2019}
S.~Grandi, M.~P. Nielsen, J.~Cambiasso, S.~Boissier, K.~D. Major, C.~Reardon,
  T.~F. Krauss, R.~F. Oulton, E.~A. Hinds, and A.~S. Clark, \enquote{{Hybrid
  plasmonic waveguide coupling of photons from a single molecule},}
  {\protect\JournalTitle{APL Photonics}} \textbf{4}, 086101 (2019).

\bibitem{Ciancico2019}
C.~Ciancico, K.~G. Sch{\"{a}}dler, S.~Pazzagli, M.~Colautti, P.~Lombardi,
  J.~Osmond, C.~Dore, A.~Mihi, A.~P. Ovvyan, W.~H. Pernice, E.~Berretti,
  A.~Lavacchi, C.~Toninelli, F.~H. Koppens, and A.~Reserbat-Plantey,
  \enquote{{Narrow Line Width Quantum Emitters in an Electron-Beam-Shaped
  Polymer},} {\protect\JournalTitle{ACS Photonics}} \textbf{6}, 3120--3125
  (2019).

\bibitem{Nicolet2007}
A.~A.~L. Nicolet, C.~Hofmann, M.~A. Kol'chenko, B.~Kozankiewicz, and M.~Orrit,
  \enquote{{Single Dibenzoterrylene Molecules in an Anthracene Crystal:
  Spectroscopy and Photophysics},} {\protect\JournalTitle{ChemPhysChem}}
  \textbf{8}, 1215--1220 (2007).

\bibitem{Trebbia2009}
J.-B. Trebbia, H.~Ruf, P.~Tamarat, and B.~Lounis, \enquote{{Efficient
  generation of near infra-red single photons from the zero-phonon line of a
  single molecule},} {\protect\JournalTitle{Optics Express}} \textbf{17}, 23986
  (2009).

\bibitem{Trebbia2010}
J.~B. Trebbia, P.~Tamarat, and B.~Lounis, \enquote{{Indistinguishable
  near-infrared single photons from an individual organic molecule},}
  {\protect\JournalTitle{Physical Review A - Atomic, Molecular, and Optical
  Physics}} \textbf{82}, 63803 (2010).

\bibitem{Major2015}
K.~D. Major, Y.-H. Lien, C.~Polisseni, S.~Grandi, K.~W. Kho, A.~S. Clark,
  J.~Hwang, and E.~A. Hinds, \enquote{{Growth of optical-quality anthracene
  crystals doped with dibenzoterrylene for controlled single photon
  production},} {\protect\JournalTitle{Review of Scientific Instruments}}
  \textbf{86}, 083106 (2015).

\bibitem{Faez2014}
S.~Faez, P.~T{\"{u}}rschmann, H.~R. Haakh, S.~G{\"{o}}tzinger, and
  V.~Sandoghdar, \enquote{{Coherent interaction of light and single molecules
  in a dielectric nanoguide},} {\protect\JournalTitle{Physical Review Letters}}
  \textbf{113}, 1--5 (2014).

\bibitem{Gmeiner2016a}
B.~Gmeiner, A.~Maser, T.~Utikal, S.~G{\"{o}}tzinger, and V.~Sandoghdar,
  \enquote{{Spectroscopy and microscopy of single molecules in nanoscopic
  channels: Spectral behavior: Vs. confinement depth},}
  {\protect\JournalTitle{Physical Chemistry Chemical Physics}} \textbf{18},
  19588--19594 (2016).

\bibitem{Polisseni2016}
C.~Polisseni, K.~D. Major, S.~Boissier, S.~Grandi, A.~S. Clark, and E.~A.
  Hinds, \enquote{{Stable, single-photon emitter in a thin organic crystal for
  application to quantum-photonic devices},} {\protect\JournalTitle{Optics
  Express}} \textbf{24}, 5615 (2016).

\bibitem{Wei2019}
S.~Wei, P.~Ren, Y.~He, P.~Zhang, and X.-W. Chen, \enquote{Single-molecule doped
  crystalline nanosheets for delicate photophysics studies and directional
  single-photon emitting devices,} {\protect\JournalTitle{arXiv:1911.04751}}
  (2019).

\bibitem{Ahangaran2019}
F.~Ahangaran, A.~H. Navarchian, and F.~Picchioni, \enquote{{Material
  encapsulation in poly(methyl methacrylate) shell: A review},}
  {\protect\JournalTitle{Journal of Applied Polymer Science}} \textbf{136},
  48039 (2019).

\bibitem{Gao2008}
Y.~Gao, S.~Reischmann, J.~Huber, T.~Hanke, R.~Bratschitsch, A.~Leitenstorfer,
  and S.~Mecking, \enquote{{Encapsulating of single quantum dots into polymer
  particles},} {\protect\JournalTitle{Colloid and Polymer Science}}
  \textbf{286}, 1329--1334 (2008).

\bibitem{Tamarat2000}
P.~Tamarat, A.~Maali, B.~Lounis, and M.~Orrit, \enquote{{Ten Years of
  Single-Molecule Spectroscopy},} {\protect\JournalTitle{Journal of Physical
  Chemistry A}} \textbf{104}, 1--16 (2000).

\bibitem{Alkan2009}
C.~Alkan, A.~Sari, A.~Karaipekli, and O.~Uzun, \enquote{{Preparation,
  characterization, and thermal properties of microencapsulated phase change
  material for thermal energy storage},} {\protect\JournalTitle{Solar Energy
  Materials and Solar Cells}} \textbf{93}, 143--147 (2009).

\bibitem{Sari2009}
A.~Sari, C.~Alkan, A.~Karaipekli, and O.~Uzun, \enquote{{Microencapsulated
  n-octacosane as phase change material for thermal energy storage},}
  {\protect\JournalTitle{Solar Energy}} \textbf{83}, 1757--1763 (2009).

\bibitem{Sari2010}
A.~Sari, C.~Alkan, and A.~Karaipekli, \enquote{{Preparation, characterization
  and thermal properties of PMMA/n-heptadecane microcapsules as novel
  solid-liquid microPCM for thermal energy storage},}
  {\protect\JournalTitle{Applied Energy}} \textbf{87}, 1529--1534 (2010).

\bibitem{Alay2011}
S.~Alay, C.~Alkan, and F.~G{\"{o}}de, \enquote{{Synthesis and characterization
  of poly(methyl methacrylate)/n-hexadecane microcapsules using different
  cross-linkers and their application to some fabrics},}
  {\protect\JournalTitle{Thermochimica Acta}} \textbf{518}, 1--8 (2011).

\bibitem{Schofield2018}
R.~C. Schofield, K.~D. Major, S.~Grandi, S.~Boissier, E.~A. Hinds, and A.~S.
  Clark, \enquote{{Efficient excitation of dye molecules for single photon
  generation},} {\protect\JournalTitle{J. Phys. Commun.}} \textbf{2}, 115027
  (2018).

\bibitem{Lombardi2018}
P.~Lombardi, A.~P. Ovvyan, S.~Pazzagli, G.~Mazzamuto, G.~Kewes, O.~Neitzke,
  N.~Gruhler, O.~Benson, W.~H. Pernice, F.~S. Cataliotti, and C.~Toninelli,
  \enquote{{Photostable Molecules on Chip: Integrated Sources of Nonclassical
  Light},} {\protect\JournalTitle{ACS Photonics}} \textbf{5}, 126--132 (2018).

\bibitem{Clear2020}
C.~Clear, R.~C. Schofield, K.~D. Major, J.~Iles-Smith, A.~S. Clark, and
  D.~P.~S. McCutcheon, \enquote{{Phonon-induced optical dephasing in single
  organic molecules},} {\protect\JournalTitle{Physical Review Letters}}
  \textbf{124}, 153602 (2020).

\bibitem{Loudon2000}
R.~Loudon, \emph{{The Quantum Theory of Light}} (Oxford University Press,
  2000), 3rd ed.

\bibitem{Rezai2018}
M.~Rezai, J.~Wrachtrup, and I.~Gerhardt, \enquote{{Coherence Properties of
  Molecular Single Photons for Quantum Networks},} {\protect\JournalTitle{Phys.
  Rev. X}} \textbf{8}, 031026 (2018).

\bibitem{Grandi2016}
S.~Grandi, K.~D. Major, C.~Polisseni, S.~Boissier, A.~S. Clark, and E.~A.
  Hinds, \enquote{{Quantum dynamics of a driven two-level molecule with
  variable dephasing},} {\protect\JournalTitle{Physical Review A}} \textbf{94}
  (2016).

\bibitem{Deperasinska2011}
I.~Deperasi{\'{n}}ska, E.~Karpiuk, M.~Banasiewicz, A.~Makarewicz, and
  B.~Kozankiewicz, \enquote{{Single dibenzoterrylene molecules in naphthalene
  and 2,3- dimethylnaphthalene crystals: Vibronic spectra},}
  {\protect\JournalTitle{Physical Chemistry Chemical Physics}} \textbf{13},
  1872--1878 (2011).

\bibitem{Knauer2017}
S.~Knauer, M.~L{\'{o}}pez-Garc{\'{i}}a, and J.~G. Rarity, \enquote{{Structured
  polymer waveguides on distributed Bragg reflector coupling to solid state
  emitter},} {\protect\JournalTitle{Journal of Optics (United Kingdom)}}
  \textbf{19}, 065203 (2017).

\bibitem{Stella2019}
U.~Stella, L.~Boarino, N.~{De Leo}, P.~Munzert, and E.~Descrovi,
  \enquote{{Enhanced Directional Light Emission Assisted by Resonant Bloch
  Surface Waves in Circular Cavities},} {\protect\JournalTitle{ACS Photonics}}
  \textbf{6}, 2073--2082 (2019).

\end{thebibliography}
\end{document}